\documentclass[prb,twocolumn,amsmath,amssymb,floatfix]{revtex4}
\usepackage{graphicx}
\newcommand{\be}{\begin{eqnarray}}
\newcommand{\ee}{\end{eqnarray}}

\newcommand{\bfr}{{\bf r}}

\newcommand{\bfp}{{\bf p}}

\newcommand{\wbe}{\begin{widetext}}
\newcommand{\wee}{\end{widetext}}
\newcommand{\oncite}{\onlinecite}

\begin{document}
\draft

\title{Confinement Induced Quantum Phase Transition and Polarization
Cooling in a Dipolar Crystal of Polar Molecules}

\author{Yi-Ya Tian and Daw-Wei Wang}

\address{Physics Department, and National Center for Theoretical Scence,
National Tsing-Hua University, Hsinchu, Taiwan, ROC}

\date{\today}

\begin{abstract}
It is well-known that the liquid properties in a strongly confined
system can be very different from their ordinary behaviors in an
extended system, due to the competition
between the thermal energy and the interaction energy.
Here we show that, in a low-dimensional self-assembled dipolar crystal,
the parabolic optical confinement potential can also strongly
affect the quantum many-body properties in the low temperature regime.
For example, by changing the confinement aspect ratio,
the bulk of the system can undergo a quantum phase transition between
a liquid state and a solid state via a nonmonotonic pattern
formation of the domain wall. Furthermore, the entropy of a trapped
dipolar crystal can be much larger than the liquid state in the
weak dipole limit, indicating an intrinsic polarization cooling
mechanism via increasing the external field.
These highly correlated confinement effects are very important
to the experimental preparation of a
self-assembled dipolar crystal using ultracold polar molecules.
\end{abstract}


\maketitle
Confinement effects are known to be very crucial to the
physical properties of a spatially confined liquid. Investigating
these extraordinary fluid properties has become
an important subject in nanoscience and technology
in recent years [{\oncite{review}}].
When the confined length scale is reduced to be comparable to the
inter-particle distance, there could be several kinds of phase transitions
observed in different systems, including
solidification on the boundary [{\oncite{phase_transition_nanofluid}}],
liquid-to-liquid phase transition [{\oncite{phase_transition_liquid}}],
and layering phenomena
[{\oncite{phase_transition_dustyplasma}}], etc.
From a microscopic point of view, all these unusual properties
result from the competition between the thermal kinetic energy
and the inter-particle interaction in a certain spatial geometry.
It is therefore reasonable to expect that the interplay between
quantum fluctuation effects, interaction energy, and the 
confinement geometry can also lead to some extraordinary 
many-body properties that are not observable in classical 
systems. Here we show that in a self-assembled low-dimensional
dipolar crystal [\oncite{Guido}], which can be formed by 
ultracold polar molecules
trapped in a magneto-optical potential, the bulk properties of the
crystal can be strongly affected by the confinement potential,
even when the system size is still much larger than the average
inter-particle distance. The thermodynamical properties are
also found qualitatively different from their behavior
in an extensive space, showing the highly correlated confinement
effect of a quantum many-body system. Different from the
observed dipolar and confinement effects in the condensate of
magnetic atoms [\oncite{Pfau}], here we mainly focus
on the dipolar crystal phase in the strongly interacting regime 
(see below), which cannot be easily achieved by magnetic atoms 
within present experimental parameter range.

We first consider polar molecules initially
prepared in the lowest rotational ground state ($L=0$) and
confined in a quasi-two-dimensional (2D) trap, where the transverse
dynamics (in $z$ axis) is frozen to the single particle ground state.
The in-plane motion of molecules, however, is weakly confined by
weaker harmonic potentials with trapping frequencies, $\omega_{x,y}$,
in the $x$ and $y$ directions respectively. When an external DC
electric field is applied along the $z$ axis, polar molecules become
polarized and have a field-dependent electric dipole moment, $D$.
In a dilute limit, the mutual interaction between these quais-2D
polar molecules is a dipolar interaction
[{\oncite{zoller_2D}}], and the system can be easily stabilized
by the strong transverse confinement potential. 
As a result, the system Hamiltonain can be written to be
\be
H=\sum_i^N\left[\frac{\bfp_i^2}{2m}
+\frac{m}{2}\left(\omega_x^2x_i^2+\omega_y^2y_i^2\right)\right]
+\frac{1}{2}\sum_{i\neq j}^N\frac{D^2}{|\bfr_i-\bfr_j|^3},
\label{H}
\ee
where $m$ is the mass and $N$ is the total number of molecules.
$\bfp_i$ and $\bfr_i=(x_i,y_i)$ are the in-plane
momentum and position operators of the $i$th particles.
There are three length scales in Eq. (\ref{H}): two
are the oscillator lengths,
$a_{{\rm osc},x(y)}\equiv\sqrt{\hbar/m\omega_{x(y)}}$, and one
is the dipolar interaction strength, $a_d\equiv mD^2/\hbar^2$.
The system properties are therefore determined by three dimensionless
parameters only: $\gamma\equiv a_d/a_{{\rm osc},x}$,
measuring the dipolar interaction strength,
$\kappa\equiv\omega_y/\omega_x=a_{{\rm osc},x}^2/a_{{\rm osc},y}^2$,
measuring the aspect ratio of the confinement potential, and particle
number $N$. Throughout this paper, we use
$a_{{\rm osc}}\equiv a_{{\rm osc},x}$ and $\omega\equiv \omega_x$ to
be the units of length and energy scales respectively.
For a typical molecule like SrO, the full polarized dipole moment is $D=8.9$
Debye and therefore $\gamma\sim 88$ for $a_{\rm osc}=1.4 \mu$m
in a typical trapping frequency,  $\omega=2\pi\times 50$ Hz.

\begin{figure}
\includegraphics[width=16cm]{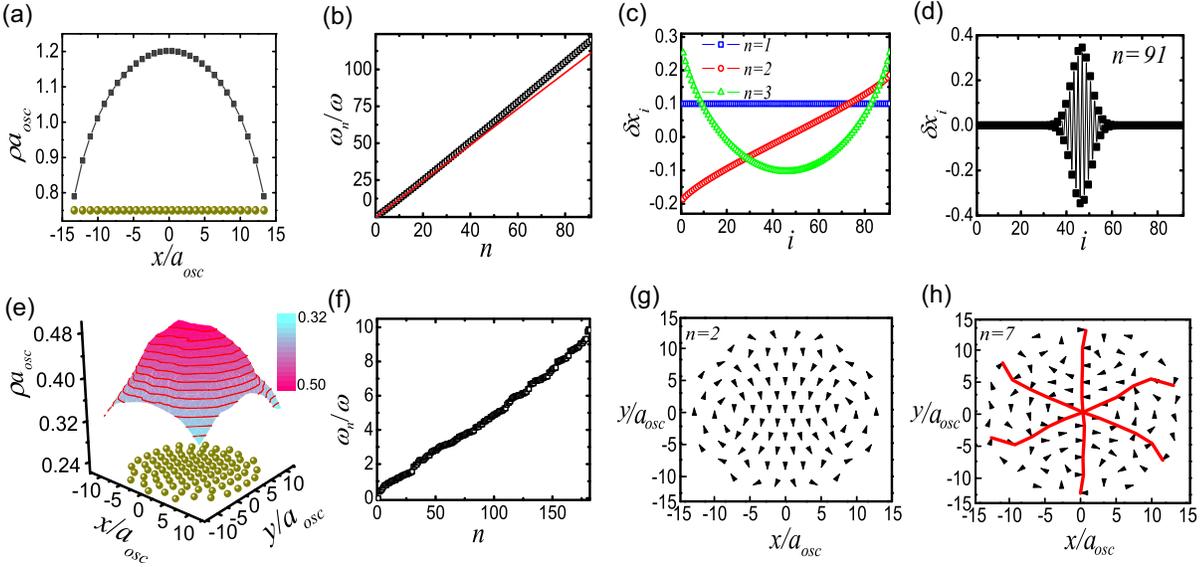}
\caption{(a) Density profile and position configuration of lattice
position for $N=30$ dipolar particles in a 1D harmonic trap with
$\gamma=17$. (b) Excitation spectrum for the same system with $N=91$.
The red line is obtained in the hydrodynamic theory of the liquid state
with $N\to\infty$: $\omega_n=\omega_x\sqrt{(3n^2-n)/2}$
[{\oncite{stringari}}]. (c) and (d) show the eigenfunction (i.e.
deviation from their equilibrium position) for the lowest
($n=1,2,3$) and the highest ($n=91$) excited states of 1D system.
(e)-(h) show the same physical quanities as (a)-(d), but 
for $N=91$ particles in a 2D isotropic harmonic trap with $\gamma=80$.  
(g) and (h) are the second and the seventh
excitation modes with vibrating direction shown by arrows.
The red lines in (h) are eye-guiding, indicating a Tachenko's mode 
as observed in the vortex lattice of a fast rotating 
condensate [{\oncite{vortex_JILA}}].
} 
\label{main}
\end{figure}

In this paper, we are interested in the regime when the dipolar
interaction is sufficiently large ($\gamma\gg 1$), so that polar
molecules can form a dipolar crystal [{\oncite{Guido}}], as shown in
Fig. \ref{main}(a) (for 1D trap, $\kappa=\infty$) and Fig. \ref{main}(e)
(for 2D isotropic trap, $\kappa=1$). Here the equilibrium position for
each molecule is calculated by using Molecular-Dynamics(MD)
simulation with a small friction in the Langevin equation
[{\oncite{dusty_plasma,MD}}]. We further calculate the phonon
spectrum by quantizing the position fluctuation of dipoles from
their equilibrium positions to the quadratic order. Within this
harmonic approximation, the obtained
spectrum (Figs. \ref{main}(b) and (f)) is independent of the dipolar
strength ($\gamma$), because the effect of tuning dipole moment can
be exactly cancelled by adjusting the inter-molecule distance
[{\oncite{stringari}}]. In Fig. \ref{main}(c)-(d) and (g)-(h), we
show some typical phonon excitation wavefunctions for these 
two systems respectively.
One easily see that the confinement potential has strongly changed the
excitation wavefunctions from simple plane waves in a uniform
system. 

\begin{figure}
\includegraphics[width=8.5cm]{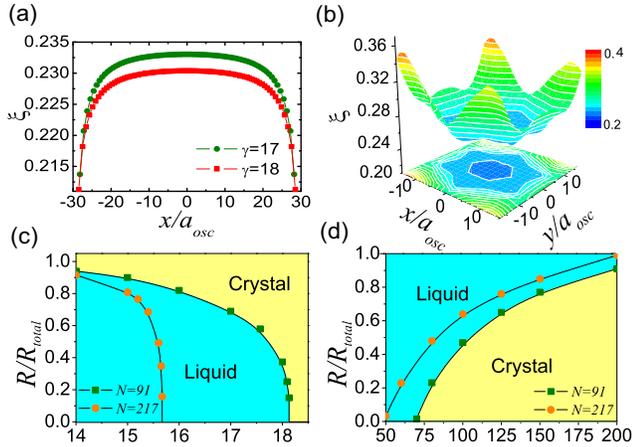}
\caption{(a) and (b): The distribution of softeness ($\xi_i$) 
for dipolar crystals with $N=91$ particles in 1D and 2D isotropic 
traps respectively. $\gamma=80$ in (b). (c) and (d): The
radii of the phase boundary, $R$, as a function of dipolar strength,
$\gamma$, for the above two systems with $N=91$. 
$R_{\rm total}$ is the radius of the whole system. Here we set
$C_L=0.23$, and phase boundary for $N=217$ are also shown for
comparison. } 
\label{main2}
\end{figure}

Following the same spirit of the Lindermann criterion,
which was first applied to the classical 3D melting problem 
[\oncite{Lindermann}] and then extended to the 
2D thermal and quantum melting 
[{\oncite{2D_melting,2D_quantum_melting,inhomogenous_melting}}],
in this paper we define that a solid
phase is melted at a certain position if its position fluctuation
is larger than an imperical ratio ($C_L$) of the
lattice constant. Such ratio in a uniform 2D dipolar crystal
has been calculated to be 0.23 from the quantum Monte Carlo 
simulation [\oncite{2D_quantum_melting,comment}].
To apply such useful criterion to a nonuniform system discussed 
in this paper, here we define the local ``softness'' of the dipolar
crystal to be: $\xi_i\equiv \Delta r_i/\bar{l}_i$, where $\Delta
r_i\equiv\sqrt{\langle(\bfr_i-\bfr_{i,0})^2\rangle}$ is the position
fluctuation of the $i$th particle (calculated from the 
wavefunction of the phonon excitation states), and 
$\bar{l}_i$ is its average distance to the nearest neighboring sites. As
a result, the $i$th particle is considered to be in a liquid
state if $\xi_i>C_L$, while it is in a solid state if $\xi_i<C_L$
[{\oncite{comment_lindermann}}]. In Fig. \ref{main2}(a) and (b), we
show the local softness ($\xi_i$) for dipolar crystal in both 1D and
2D isotropic potentials. We find that in 1D system the softness
distribution is convex with its maximum value in center of trap,
i.e. the bulk of the 1D dipolar crystal is softer than the edge, in
contradiction to a naive guess from a local density approximation,
which predicts the position of higher particle density should be 
more solid-like due to the $r^{-3}$ dipolar interaction. 
This striking results originate from the confinement
effect on the quantum fluctuation of lattice points: the system
ground state wavefunction has equal contribution from each
eigenmodes (note that quantum zero point energy is $E_Q=\sum_n
\frac{1}{2}\hbar\omega_n$), while higher energy modes prefer to
generate much larger position fluctuation in the center of the 1D
trap (see Fig. \ref{main} (c) and (d)). In the 2D isotropic trap, on
the other hand, the softeness becomes concave, because now the high
energy eigenmodes can still have large amplitude at the edge due to
the additional degree of freedom in the azimuthal direction of the
trap (see Fig. \ref{main}(g) and (h)). In Fig. \ref{main2}(c) and (d), we
further show the phase boundary (domain wall) between the liquid
state and the solid(crystal) state as a function of dipolar strength
($\gamma$): the 1D system has a central domain of liquid state as
long as $\gamma$ is smaller than a critical value, $\gamma_{\rm
1D}^\ast$, above which the whole system becomes a well-defined solid
crystal. On the other hand, the center of the 2D isotropic trap
starts to be crystalized only when $\gamma$ is larger than another
critical value, $\gamma_{\rm 2D}^\ast$. Such qualitative difference
in these two systems are clear evidence of the highly correlated
confinement effects on the quantum fluctuations. Both $\gamma_{\rm
1D}^\ast$ and $\gamma_{\rm 2D}^\ast$ decrease as the number of
particles increases.

\begin{figure}
\includegraphics[width=9cm]{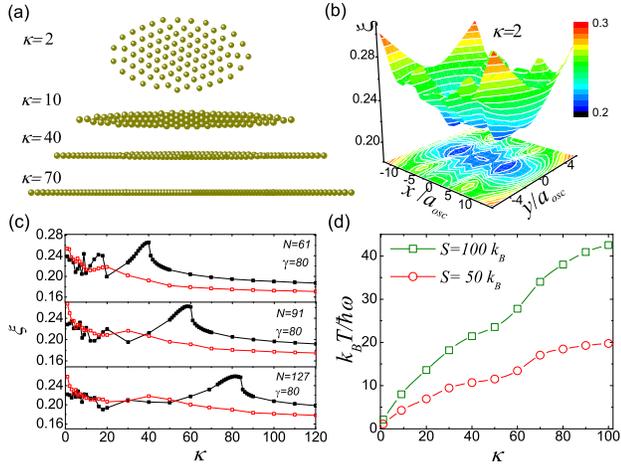}
\caption{
(a) Position configuration of the dipolar crystal ($N=91$) in elliptic traps
of different aspect ratios, $\kappa$. (b) Distribution of the softness 
for $\kappa=2$. (c) Local softness
in the center (filled squares) and at the end (open squares) for systems of
different numbers of particles ($N$).
(d) The system temperature changes during an adiabatic 
compression from a 2D isotropic trap ($\kappa=1$) to 1D trap 
($\kappa\gg 1$) with $N=91$.
The confinement frequency in the $x$ direction is fixed, and all results of
are calculated with $\gamma=80$.
}
\label{elliptic}
\end{figure}

In Fig. \ref{elliptic}, we show the confinement effects in
different trap aspect ratio with a fixed $\omega_x$: the anisotropic trapping potential
generates two local minimum of the softness (see Fig. \ref{elliptic}(b))
along the $x$ axis. Within a proper parameter range, these two local
minimum of softness can become the centers of two solid
island(domain) embedded in a dipolar liquid.
In Fig. \ref{elliptic}(c), we show how the local softness,
in the edge (red lines) and in the center (black lines) of the
trap, changes as function of the aspect ratio, $\kappa$.
We find that although the former decreases monotonically
as expected, the later can have several re-entrant effects and
can be even larger (i.e. softer) than its value in the isotropic trap
($\kappa=1$) without compression.
Increasing number of particles does not change these
nonmonotonic bulk properties (say domain walls and re-entrant behavior),
showing a significant confinement
effects even when the system size is much larger than inter-particle
distance. Similar nonmonotonic/reentrant behavior are also observed in
a system of classical 2D melting [{\oncite{inhomogenous_melting}}]
but much more particles and quantum fluctuation effects are
considered here.

The nontrivial confinement effect can be also observed in the finite(but low)
temperature regime by investigating the system entropy ($S$),
which can be assumed to be conserved during an adiabatic
manipulation of the system parameters. The total
entropy can be calculated from [{\oncite{KersonHuang}}]
$S=\int_0^TdT' {C_v(T')}/{T'}$, where $C_v(T)=\partial E(T)/\partial T$
is the specific heat. The total energy, $E(T)$, can be easily
calculted from the phonon excitation spectrum: $E(T)=E_C +
E_Q+\sum_n\frac{\hbar\omega_n}{e^{\hbar\omega_n/k_BT}-1}$,
where $E_C$ is classical potential energy and $k_B$ is Boltzman constant.
In Fig. \ref{elliptic}(d), we show how the system temperature changes
as a function of the confinement aspect ratio, $\kappa$, by
keeping the total entropy a constant during the adiabatic process:
the temperature increases sublinearly as the system is
compressed and eventually becomes saturated in the limit of pure
1D system (say $\kappa> 100$ for $N=91$).
Within the harmonic expansion approximation for the phonon excitations,
the calculate system entropy is independent of the dipolar 
strength, $\gamma$.

\begin{figure}
\includegraphics[width=7cm]{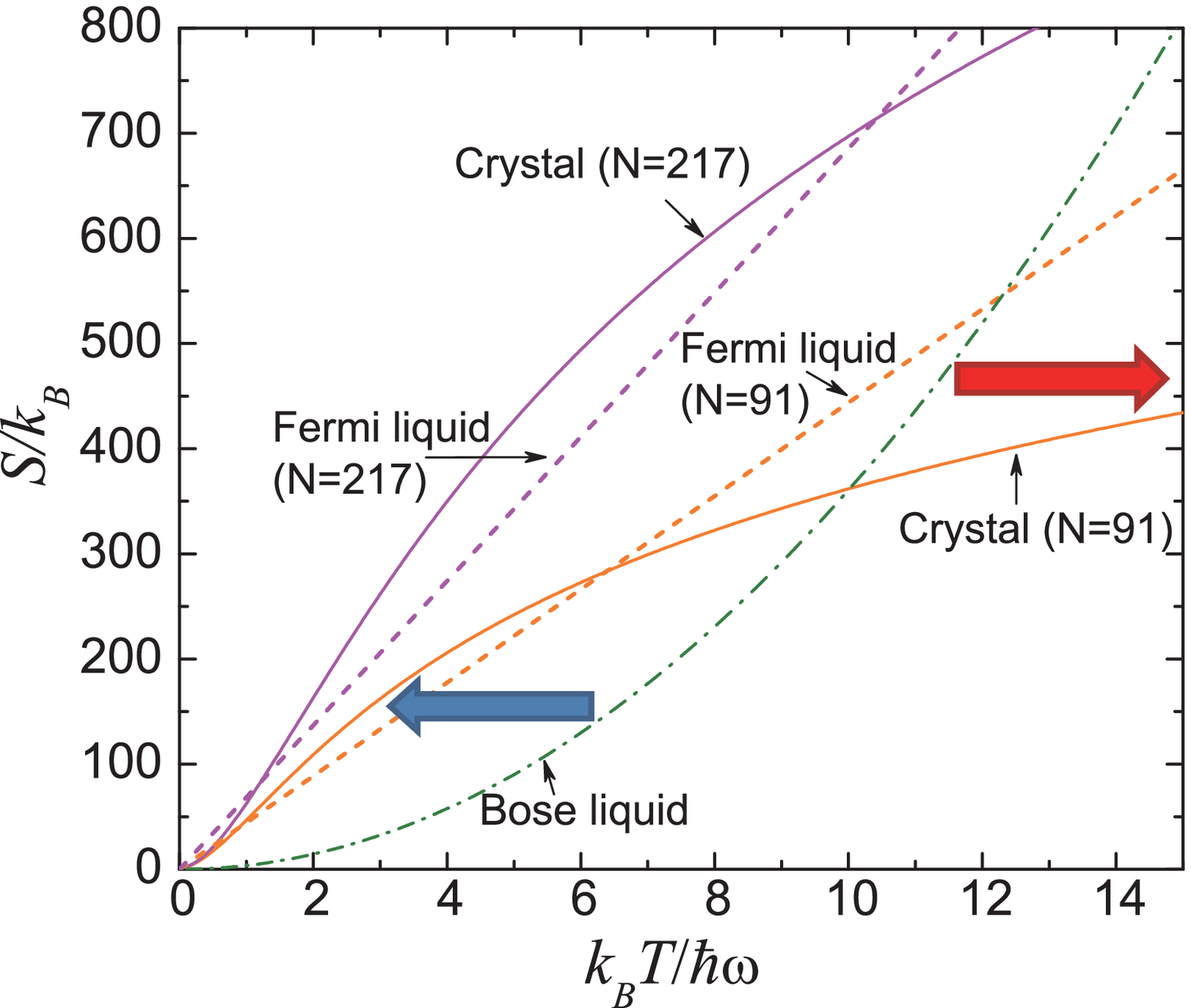}
\caption{Total entropy as a function of system temperature in 
a 2D isotropic confinement potential.
Results for Bose liquid (dash-dotted lines), Fermi liquid (dotted lines)
and crystal phase (solid lines) are shown together. Results for different
numbers of particles are also shown for comparison. The blue leftward 
arrow indicates the intrinsic cooling process for a Bose liquid, while
the read rightward arrow indicates the intrinsic heating process.
}
\label{entropy}
\end{figure}

From experimental point of view, one of the most important question
is how the system temperature changes when a dipolar crystal is
formed by increasing the electric field.
In the zero dipole moment limit (i.e. zero external field),
only $s$-wave scattering exists between bosonic
molecules, while it is almost noninteracting for single component fermionic
molecules. For a Bose liquid in a 2D isotropic trap, we can
apply the local density approximation to
calculate the system entropy [{\oncite{stringari2}}], and obtain
the leading order temperature dependence:
$
S_{\rm BL}/k_B=
\alpha\frac{\pi^2}{\sqrt{2}}\left(\frac{k_BT}{\hbar\omega}\right)^2
$
with $\alpha\sim 5.7\times 10^{-3}$ being obtained from direct
numerical calculation. This result is independent of the $s$-wave
scattering length and particle number due to the unique geometry of 2D
isotropic harmonic trap. Similar to the 3D case [{\oncite{stringari2}}],
it applies to condensate as well as to the normal state, because the
dominate contribution of entropy always comes from the single particle
excitation of normal liquid.
For the noninteracting fermionic molecules, we can also calculate the
system entropy easily from the temperature dependence of
the total energy in a 2D harmonic trap [{\oncite{bec}}], and obtain
$
S_{\rm FL}/k_B=\frac{\pi^2\sqrt{2N}}{3}
\left(\frac{k_BT}{\hbar\omega}\right).
$
In Fig. \ref{entropy} we show the calculated entropies
of Fermi liquid (dotted lines), Bose liquid (dash-dotted line)
and dipolar lattice (solid lines) in a 2D isotropic trap.
Results for both $N=91$ and $N=217$ are shown together for comparison.
It is easy to see that if bosonic polar molecules are initially prepared
at zero field and in a sufficiently low temperature ($<T_{\rm boson}^\ast$),
the system temperature will decrease (intrinsic cooling,
blue leftward arrow) as the external field is increased to derive
the system toward a dipolar crystal adiabatically.
On the other hand, if the system is prepared in a rather high
initial temperature ($>T_{\rm boson}^\ast$), the system
temperature will increases greatly as the dipole moment increases
(intrinsic heating, the red rightward arrow). The critical
temperature, $T^\ast_{\rm boson}\sim 10\hbar\omega/k_B$ for $N=91$,
but becomes about $16\hbar\omega/k_B$ for $N=217$. This shows
that the polarization cooling here is due to the
many-body effects of polar molecules, completely different from
the demagnetization cooling process in solid state systems or
in magnetic dipolar atoms [{\oncite{depolarization}}].
More precisely, due to the presence of a harmonic confinement,
the average distance between dipoles increases (i.e. the average
density decreases) as the dipole moment is enhanced, and therefore
the system temperature can be reduced via transferring the electric
field energy to the confinement potential energy (rather than to the
kinetic energy) in the limit of strong dipolar crystal.
Such intrinsic cooling mechanism is totally different from what
is expected in a uniform system, where the entropy of a crystal
should be always smaller than a liquid state at the same density
[{\oncite{He3}}].
Similar intrinsic cooling mechanism can be also
observed in fermionic polar molecules and/or in 1D trapped system.

In summary, we have shown several important confinement
effects on a self-assembled dipolar lattice formed by
ultracold polar molecules.
Changing confinement aspect ratio can induce a quantum
phase transition between a bulk liquid state in 1D trap to a bulk
solid crystal state in 2D isotropic trap.
We further find an intrinsic polarization cooling mechanism during
the formation of dipolar crystal. Our results can be applied
to the experimental preparation of a dipolar crystal
of ultracold polar molecules and are also important to the
understanding of the confinement effects in a quantum many-body system.

We thank fruitful discussion with G. Pupillo, P. Zoller, B.I. Halperin, 
and Lin I. This work is supported by National Science Concil.





\begin{thebibliography}{99}
\bibitem{comment}
It is believed that the classical melting of a 2D crystal is
better described by the
seminal Kosterlitz-Thouless-Haperin-Nelson-Young (KTHNY) theory
[{\oncite{KTHNY}}], which
suggests a two-stage process mediated by the pairwise
creation of topological defects.
However, since we are interested in the quantum phase
transition in an inhomogeneous system at low temperature,
topological defects should not play important role here.
\bibitem{comment_lindermann}
We note that although there are slightly different definition and
values of Lindermann constant, $C_L$, in the literature, but it
will not affect any of our conclusion qualitatively.
\bibitem{He3}
The only known exception is probably the solid state of $^3$He, which
has additional spin degree of freedom to reserve extra entropy compared
to its liquid state.
\end{thebibliography}
\end{document}